\documentclass[twocolumn,showpacs,preprintnumbers,amsmath,amssymb]{revtex4}

\usepackage{graphicx}

\begin{document}

\title
{The static and dynamic conductivity of warm dense Aluminum and Gold
calculated within a density functional approach.
}

\author
{
 M.W.C. Dharma-wardana}
\affiliation{
Institute of Microstructural Sciences,
National Research Council of Canada, Ottawa,Canada. K1A 0R6
}
\email[Email address:\ ]{chandre.dharma-wardana@nrc.ca}
\date{\today}

\begin{abstract}
The static resistivity of dense Al and Au plsmas are
calculated where all the needed inputs are obtained
from density functional theory (DFT). This is used
as input for a study
of the dynamic conductivity.
 These calculations 
 involve a self-consistent determination of  (i) 
the equation of state (EOS) and the ionization balance,
 (ii) evaluation of the
ion-ion, and ion-electron  pair-distribution functions.
 (iii) Determination
of the scattering amplitudes, and finally the conductivity.
 We present data for 
for the static resistivity of 
Al for compressions 0.1-2.0, and in the 
 the temperature range
$T$ = 0.1 - 10 eV.  
Results for Au in the same
temperature range and for compressions 0.1-1.0 is also given.
In determining the dynamic conductivity for a range of frequencies
consistent with standard laser probes, a knowledge of the
electronic eigenstates and occupancies of Al- or Au plasma  
 become necessary. They are calculated using a neutral-pseudoatom model.
We examine a number of first-principles approaches
to the optical conductivity, including 
 many-body perturbation theory,  
molecular-dynamics evaluations, and simplified
time-dependent DFT. The modification to the
Drude conductivity that arises from the presence of
shallow bound states in typical Al-plasmas is examined and
numerical results are given at the level of the Fermi Golden rule
and an approximate form of time-dependent DFT.
\end{abstract}
%
\maketitle
\section{Introduction}
 The static conductivity of matter in the plasma state can 
be calculated with some confidence, at least for a number of
``simple'' plasmas, using an entirely first-principles approach
\cite{per87,eos95}. Comparisons with experiments are
available for a wide range of conditions,
 at least for Al plasmas
\cite{benageR}.
 Another  transport property of great interest
is  the frequency-dependent
conductivity $\sigma(\omega)$.
 Optical probes provide a convenient diagnostic tool for this
plasma conductivity, and common 
laser-probe wavelengths, e.g., 300-800 nm, become the window
of interest. In fact, many experiments on laser-plasma interactions
related to inertial fusion have concentrated on frequency tripled
(with the fundamental $\omega$ at 1$\mu$m from glass lasers)
light,  $3\omega$, 351nm, while there is also current interest
in the $2\omega$ (527 nm) regime as an option for 
indirect-drive ignition\cite{niemann}.
Normally, if the probe frequency is less than the
plasma frequency, the probe photons fail to enter the material
and only weakly ionized systems can be accessed. However, the
recent development of extremely thin ("nanoscale")
 plasma-slab techniques\cite{ng}, known as idealized-slab plasmas\cite{ngcdw}
has made it possible to study dense plasmas with low-frqeuency probes,
both in transmission and in reflectivity.

Density-functional calculations are of two types.
The first type depends heavily on quantum MD 
(QMD) simulations, e.g, Car and Parrinello
 treat {\it only the electrons} via Kohn-Sham theory,
 while the
ion subsystem, explicitly represented by a convenient number $N$ 
(of the order of $\sim$ 32-256)  of ions, is made to evolve in time
and an average over millions of configurations is taken. In full
Quantum Monte Carlo (QMC) simulations\cite{ceperley81,kwon}, even
 Kohn-Sham theory is not
used and hence full QMC is not practical for typical plasma problems.
The second type of DFT is typified by
 our approach where both {\it electrons and ions} are treated by DFT, so
that both subsystems are described by two
 coupled ``single-electron'' and ``single-ion'', Hartree-like
 Kohn-Sham
equations. The many-body effects (i.e, many-electron and many-ion
effects) are included through the exchange-correlation or correlation
potentials. This allows an enormous simplification 
in the numerical work involved.
These equations may be further reduced to  extremely simpilified 
Thomas-Fermi approaches yielding approximate results for
plasma properties which are within an order of magnitude of the more
refined results, even in unfavourable cases\cite{leemore}.

The general optical conductivity problem
is essentially the same as that of opacity calculations for 
warm dense matter\cite{gri}.
However, here we approach it from the 
static-conductivity ($\omega=0$) regime, and take account of
free-free processes as well as
shallow bound states which may exist.
The
static-conductivity provides an 
evaluation of the dynamic conductivity near $\omega$ = 0
via the basic Drude formula which assumes
a constant relaxation time $\tau$,  taken to be the {\it static} collision
time $\tau(0)$. Hence our task can be stated as follows:
\begin{enumerate}
\item
 {Calculation of the static resistivity. This involves
the following steps:}
      \begin{enumerate}
 \item
	{Calculate the Kohn-Sham atom in an electron gas of density $n$.}
 \item
 	{Use the Kohn-Sham results to form pseudopotentials $V_{ie}$ and 
        scattering cross sections at the given density $\rho$ and
        temperature $T$.}
 \item
	{Use the $V_{ie}$ to form pair potentials and pair-distribution
     functions. Here the Kohn-Sham equation for the ion-subsystem
     can be approximated by some form of the
     	hyper-netted-chain equation (HNC) with bridge terms.}
 \item
	{Calculate the EOS, ionization balance etc., to obtain
	effective ionic charges $\bar{Z}$, electron density
	 $n=\rho/\bar{Z}$etc., and
        self-consistently, repeating from item (a).}
 \item
	{Calculate the static resistivity and the static
         relaxation time $\tau(0)$
         using, e.g, using a  Ziman-type
        formula valid for strong coupling and finite $T$.}
       \end{enumerate}
 \item
 {Use the energy-level structure of the Kohn-Sham atom 
 for the $\rho$
 and 
 $T$ regimes of interest to set up  bound-bound, or bound-free
 processes which fall within the range of frequency $\omega$  considered.
 Corrections may be necessary since the Kohn-Sham theory does not 
 provide good excitation energies.
 }

\item
 {Construct the dynamic conductivity $\sigma^0(\omega)$.  
This yields results equivalent to the
 Fermi golden rule. 
}
\item
 { Extend the calculation of $\sigma(\omega)$
 using time-dependent density-functional
 theory\cite{gri}.Include the coupling of
  electronic transitions to ion-dynamics.}
\end{enumerate}

While some aspects of this program can be carried out, 
it is fair to say that a complete, consistent theory
of the transport properties of such many-body systems as posed  by warm dense
matter,  or indeed, even the static properties as embodied
 in the equation of state {\it for some regimes of density and temperature}
are open to debate, even for well studied systems like
 aluminum\cite{eosb},  and hydrogen\cite{hyd}.
The objective of the present study is to calculate the dynamic conductivity
of Al and Au plasmas for a number of compressions and densities where
the Drude theory may need correction due to the presence of
shallow bound states. Such boundstates ionize when the compression is
changed, and produce distinctive changes in transport properties.
Sharply rising static resistivities
under a change of compression is a common feature of some of the theoretical
calculations shown here. However, the establishment of a genuine phase
 transition requires more care\cite{eos95}. The  $\sigma(\omega)$ of expanded
liquid metals and plasmas show\cite{bhat} effects arising from clustering
and excitonic effects, as the metal-insulator transition is approached.
These excitonic effects are not important in dense systems.
While Al has been an object of extensive study,
recent experiments in the warm dense matter regime 
have focused on gold targets\cite{japmore,aung}. Here we present
numerical results for Al and Au  for
several compressions and temperatures.
\section{static resistivity}
We use atomic units (Hartree=1 a.u., the Bohr
radius $a_0$ = 1, with $|e|=\hbar=m_e=1$). 
The atomic unit of resistivity, given by $\hbar a_0/e^2$
has a value of 21.74 $\mu\Omega$cm.
If  the equilibriation of the electron distribution
perturbed by the applied electric field
is governed by a  relaxation time $\tau$, the conductivity
$\sigma$ is given as:
\begin{equation}
\sigma=\frac{\omega_p^2}{4\pi}\,\tau
\end{equation}
A mean free path $l_{mfp}$ = $<v>/\tau$,
where $<v>$ is some characteristic mean velocity, is often
introduced. If electrons
were classical {\it point} particles,
 then it is evident that $l_{mpf}$ cannot be 
smaller than the mean separation between collision centers. This is
sometimes called the Joffe-Regel-Mott rule, and holds well in
many semiconductors. However, electrons
are quantum particles (wave packets) 
with an extension of the order of the thermal de Broglie
length. 
Further,  $\tau$ depends on the electron momentum $k$. Thus
there are examples where
$l_{mpf}$, obtained  by some averaging process, is in fact smaller than
some estimated ``mean-ion separation''. 
Although the concept of the mean free path is implicit in the
Boltzmann-equation approach, this is not necessary
 in the Dyson equation which replaces
the Boltzmann equation in the quantum case. 
If the mean free path is large, the ``particle picture'' of the electron
applies, while if this is comparable to the lattice parameter,
then we are in the diffraction limit and the wave picture
must be used. The Dyson equation, valid at both extremes, describes the
one-particle propagator which is closely related to the distribution function
appearing in the Boltzmann equation. 
The derivation of a transport coefficient
from the Dyson equation  leads us to the current-current correlation function.
This is closely related to various two-body distributions and collision 
kernels found even in classical kinetic models.
 Unfortunately, although formal expressions
can be written down, their evaluation using Green's functions or related 
methods becomes impractical, especially if bound states are present.
In effect, Green's-function methods can be pushed
to, say, second order in the screened interactions. Such
an approach is sufficient if there are no boundstates associated with
the scattering potential. Attempts to go further
rapidly
become intractable and useless.

Our approach is to use a variety of techniques 
and replace the ion-electron interactions by
suitably constructed pseudopotentials, or use scattering cross sections
calculated from  phase shifts. This requires a
fairly sophisticated non-perturbative  description of the
ion-ion and ion-electron correlations in the plasma.
\subsection{Description of the plasma using the Kohn-Sham equations}
We begin with the bare nuclei and construct the
electronic and ionic structure of the plasma.
The interaction of the nucleus with the electron fluid is a highly nonlinear
process and attempting to treat it using perturbation theory 
is  unfruitful. Hence we use the
 Kohn-Sham technique, and construct the non-linear charge density
 around the nucleus.  The nucleus, together with its
charge cloud of bound 
states and continuum-electron states
constitute a neutral object. This neutral object is called
the {\it neutral pseudo-atom} (NPA), following the
usage of e.g., Ziman and Dagens\cite{dagens}.
 Thus an  important result of the
Kohn-Sham procedure is the charge-density ``pileup'', $n(r)$,
around the nucleus that essentially screens the nucleus.
A part of this arises from the free electrons and is denoted by $n_f(r)$. 
This $n(r)$ and $n_f(r)$  depend on the mean density ${\bar n}$ of the electron fluid,
 temperature $T$,
 and the nuclear charge $Z$.
 The Kohn-Sham procedure leading to the
 NPA provides the phase shifts suffered by the continuum electrons when
 they scatter from the nuclei. These are used
for constructing the scattering cross sections 
(or pseudopotentials) which describe the electron-ion interaction.
The pseudopotential has an effective ionic charge $\bar{Z}$ and 
it behaves as  $-\bar{Z}/r$ for large $r$. The rapid oscillations of the
potential near the nucleus is replaced by a weak, smooth core region
as the ``valence'' electrons do not really penetrate the ion-core.
The pseudpotentials used in many of the solid-state or molecular
 code packages\cite{codes}
have the necessary transferability and could be quite useful. However,
they assume that the pseudopotentials
 would be used within a Schrodinger or Kohn-Sham type
procedure rather than in a linear-response scheme, and hence
they  cannot be
directly used within a Ziman-type formula.

Sometimes, instead of using the all-electron Kohn-Sham equation or
using a suitably constructed pseudopotential,
the electron-ion interaction is
replaced by a Yukawa-type interaction (effectively, a Debye-screened
interaction). The electronic structure is calculated for such a
static-screened nucleus. This procedure is not justifiable since
the energies of bound-state electrons correspond to very high frequencies
at which there is no screening. Further, the orthogonality of the
continuum eigenstates and the bound states ensures that there is very little
penetration of the free electrons into the bound-electron region.
Consequently, there is very little screening of the inner bound states,
where as the Yukawa potential screens the inner bound states as well.
Thus
calculations using a Debye-like  potential in warm dense matter
is likely to be
 incorrect irrespective of
whether  the Born approximation, or  a T-matrix, etc., were used.

 In the case of Al and also Au, it is
possible to construct, in many situations, a soft
pseudopotential $V_{ie}(q)$
which is
weak in the sense that it is possible to recreate the non-linear
electron-density ``pileup'' $n_f(r)$, obtained via the Kohn-Sham equation, by
within {\it linear} response theory. 
That is, we {\t define} the $V_{ie}(q)$ such that
\begin{equation}
n_f(r)=-V_{ie}(q)\chi(q)
\end{equation}
Here $\chi(q)$ is the electron linear-response function. 
Unlike the transferable pseuodpotentials used
in {it ab initio} packages, this $V_{ie}(q)$ is specific to the
chosen $Z$, $\bar{n}$,  $T$, and
the atomic number $Z$. It is often convenient to write the
pseudopotential in the form 
\begin{equation}
V_{ie}(q)=\bar{Z}M_qv(q), \; v_q=4\pi/q^2
\end{equation}
where $v_q$ is the bare Coulomb potential and $M_q$ is a
form factor. Only a local pseudopotential is used, and this is
quite adequate for an analysis of the experimental data
currently available.
Since the pseudopotential is weak by construction, the ion-ion
pair potential can be taken to be
\begin{equation}
U_{ii}(q)=4\pi\bar{Z}^2/q^2 + |V_{ie}(q)|^2\chi(q)
\end{equation}
Given $U_{ii}(q)$,
the ion-ion
distribution function $g_{ii}(r)$ and the structure factor $S(k)$ 
can be calculated using the
Hyper-netted-chain (HNC) equation or its extension where a bridge
term is included.  We note (see below) that the HNC equation
 (or its extension) is in fact the Kohn-Sham equation for classical
particles ( here, Al$^{z+}$ or Au$^{z+}$ ions ) for  certain choices of
 the ion-ion correlation
potential (there is no exchange potential because the ions are classical
particles).

To summarize, by using the Kohn-Sham procedures, we have thus obtained
the ion-electron pseudopotential $V_{ie}(q)$, the structure factor
$S_{ii}(q)$, and the charge density $n_{ie}(r)$, and an effective ionic
charge $\bar{Z}$ which enters into the pseudopotential. The
phase shifts have been used to construct a
 scattering cross section\cite{per87} which
may be used instead of the pseudopotential.
Hence we have a completely self-consistent procedure for obtaining
all the relevant quantities starting from the nuclear charge of the element.
The numerical codes for carrying out these procedures 
are available via the internet, to any interested
researcher\cite{web}.

\subsection{Is this a ``one-center'' approach ?}
To answer this question, we
 consider the density-functional theory of a two component system
 consisting of electrons (density profile $n(r)$ ), and ions, with a
 density profile $\rho(r)$, with respect to an ion positioned at the
  origin\cite{ilciacco}.
 Then the Hohenberg-Kohn-Mermin theorem states that the free energy
 $F[n(r),\rho(r)]$ is a functional of the density distributions such that:
  \begin{eqnarray}
    \partial F[n(r),n(r)]/\partial n(r)&=&0\\
  \partial F[n(r),\rho(r)]/\partial \rho(r)&=&0.
  \end{eqnarray}
  The first of these equations leads to an effective
   {\it single-electron} equation,
  viz., the Kohn-Sham equation
  where the effective potential contains an exchange-correlation potential
  which takes account of many-electron effects. The second equation also leads
  to a Kohn-Sham equation which is a classical equation for a {\it single ion}.
   This also contains
  an ion-ion correlation potential 
  which brings in the effects of the multi-centered
  system. It can be seen that this classical Kohn-Sham
   equation reduces to the
  HNC equation for a certain choice of the ion-correlation potential.
  Further, for ``simple'' metallic plasmas like Al where the pseuopotential is
  weak, this
  scheme relates closely to pair-potential based 
  liquid-metal theory. Such a simplification does not hold,
   for e.g., for hydrogen
  plasmas\cite{hyd}. Since ion-ion, ion-electron and electron-electron
  correlations are included in the theory, it is {\it not} a 
   single-ion model of the
  plasma. It is firmly rooted in a many-electron, many-ion DFT approach 
  which does {\it not} invoke the Born-Oppenheimer approximation.
   Since the theory
  is explicitly based on distribution functions, it is manifestly non-local
 and can be
  easily implemented to be free of electron self-interaction errors.
  Our approach may be contrasted with the Car-Parrinello (CP)
   approach where DFT
   is
  used only for the electrons, while the ions are explicitly and individually
   treated by classical molecular dynamics.
   CP avoids the need for an ion-correlation potential, but demands
   a much larger computational effort. Some of these efforts, based on MD
   methods (e.g, Ref.~\cite{kwon,surh}) have confirmed results obtained
   by the numerically simpler methods that we have used. 
\subsection{Extended Ziman formula for strongly-coupled
 electrons and ions.}
 The Ziman formula is an application of the Boltzmann equation to
liquid metals.
It was extended to finite
 temperatures by a number of authors.\cite{geoffry}
The crux of the problem is the evaluation of the
collision rate. Compared to some methods well known in plasma theory
 (e.g, Lenard-Balescu) the collision rate is easily evaluated using the
``Fermi golden rule''.  In this section we briefly
recapitulate Ziman theory
in the language of the Fermi golden tule.
In the relaxation-time approximation we
assume that the perturbed Fermi distribution $f(k)$ for electrons
 with momentum $\vec{k}$ relaxes towards the
equilibrium distribution $f_0(k)$ according to the equation:
\begin{equation}
\label{tau}
-\frac{\partial f}{\partial t}\big|_{col}=\frac{f(k)-f_0(k)}{\tau(k)}
\end{equation}
 Considering an electron scattered elastically from state $\vec{k}$
to state $\vec{k'}$, with $|k|=|k'|$, $\epsilon_k=\epsilon_{k'}$, the net
scattering rate is the difference of the
two processes  $(\vec{k}\to \vec{k'}) - (\vec{k'}\to \vec{k})$.
The initial and final densities of states for the $\vec{k}\to \vec{k'}$
process is $f(k)$ and $[1-f(k')]$. Hence the Fermi Golden rule gives
\begin{eqnarray*}
R(k\to k') &=& (2\pi/\hbar)\sum|T_{kk'}|^2
\delta(\epsilon_k-\epsilon_{k'})f(k)(1-f(k')\\
R(k'\to k) &=& \mbox{ permutation of $k$ with $k'$ etc.,}
\end{eqnarray*}
Since $\epsilon_k=\epsilon_{k'}$, this involves only an angular
integration and $|k|=|k'|$. Since the energy is not changed, the
static  resistivity
arises {\it purely from momentum randomization}.
 The change in momentum is
$q^2=2k(1-cos(\theta))$.
Here $\theta$ is the angle between $\vec{k'}$ and $\vec{k}$. 
This $(1-cos(\theta))$ term  does not
 appear in the usual relaxation time which is the time between scattering
events. Using these rates in Eq.~\ref{tau},
 we obtain a result
for the {\it inverse} of the relaxation time. 
\begin{equation}
 \frac{1}{\tau(k)}=2\pi\sum\delta(\epsilon_k-\epsilon_k')|T_{kk'}|^2
(1-cos(\theta))
\end{equation}
Here the sum merely indicates the integration over $\theta$.
The ``$T$-matrix'' appearing here describes the scattering of an electron
by the whole ion-distribution (i.e, not just one ion).
Given a set of ions at instantaneous positions
$\vec{R_I}$, then the interaction of an electron at $\vec{r}$
with the whole distribution  is of the form:
\begin{equation}
V(r)=\sum_I V_{ie}(\vec{r}-\vec{R}_i)
\end{equation}
The matrix element between the initial state
$\vec{k}$ and the final state $\vec{k'}$,
 with $\vec{q}$ = $\vec{k'} - \vec{k}$ is:
\begin{equation}
V(q)=\sum_I \bar{Z}M_q v_q e^{iq\cdot\vec{R_I}}.
\end{equation}
Note that
\begin{eqnarray}
\rho_q&=&\sum_I exp(i\vec{q}\cdot\vec{R_I})\\
<\rho_q\rho_{-q}>&=&N_iS_{ii}(q).
\end{eqnarray}
Thus the ion-ion structure factor $S(q)$ and the single-ion
scattering cross section (or the pseudopotential) 
from a single ion combine to give the full scattering $T$-matrix.
  The dependence on the
structure factor becomes negligible for $T> 10$ eV. The
individual scattering cross section can be replaced by a single-center
$T$-matrix (to be denoted by $t_{kk'}$) obtained from the phase shifts of the
NPA calculation for a single nucleus. The $S(q)$ is also obtained from the
 pair-potential constructed
from the same NPA calculation. In Ref.~\cite{per87} we showed how
to  avoid the calculation of the $S(q)$ by directly computing the
scattering cross section from the {\it whole ion distribution}.  That is,
$T_{kk'}$ s not factored into single-center $t_{kk'}$ and the associated
structure factor. Such an approach  is needed
 for {\it strongly interacting systems} where such a
factorization may not be valid. In this context we 
note that the resistivities calculated by us using the "single-center" model
 (ie., $t_{kk'}S(q))$, and the full ion-distribution model ( $T_{kk'}$
for strong-scattering) for H-plasmas were independently confirmed by
the quantum Monte Carlo simulations of Kwon et al\cite{kwon}.
However, if the pseudopotential is weak, the $S(k)$ and the single-center
$t(k)$ may be used.

Given the inverse relaxation time $1/\tau(k)$ for an electron of
momentum $k$, or equivalently, $1/\tau(\epsilon)$ for the
energy, $\epsilon = k^2/2$, we need to average this
 over all electron energies to
obtain a resistivity or a conductivity. The averaging used in the Ziman
formula leads to a {\it resistivity}, while a direct application
of the Boltzmann equation would lead to a conductivity.
Thus,
\begin{equation}
\sigma = \frac{\omega^2_p}{4\pi}<\tau(\epsilon)>, \;\;
  R =\frac{4\pi}{\omega^2_p}<1/\tau(\epsilon)>
\end{equation}

The Boltzmann equation shows that the averaging relevant
 to the conductivity calculation is
such that
\begin{equation}
\sigma=\frac{\omega^2_p}{4\pi}\frac{2}{3n_e}
\int \frac{d \epsilon}{\pi^2}(\surd{2}
\epsilon^{3/2})\frac{-\partial f_0(\epsilon)}{\partial \epsilon}\tau(\epsilon)
\end{equation}
On the other hand, the averaging over the $1/\tau(\epsilon)$ used in the
extended Ziman formula for the resistivity is somewhat different.
\begin{equation}
< 1/\tau> = -\int_0^\infty d\epsilon g(\epsilon)
\frac{\partial f(\epsilon)}{\partial \epsilon} \tau(\epsilon)^{-1}
\end{equation}
Here $g(\epsilon)$ is a density-of states factor which is unity for
free non-interacting electrons. It should be constructed from the
 phase shifts in a
strongly scattering environment.

 The initial assumption, Eq.~\ref{tau},
was that the  modified
distribution was defined via a relaxation time. A more complete
approach is to represent the modified part $\delta f(k)$ as a
series expansion in a set of suitably constructed orthogonal polynomials, 
and obtain a variational solution. The set of polynomials appropriate
for degenerate (and partially degenerate) electrons has been discussed
by Allen.\cite{allenpoly}. In the classical limit, such polynomials are
the well known Sonine polynomials. The usual relaxation-type approach is
equivalent to a single-polynomial solution. This is
adequate for dense Al-plasmas, and for the range $0 < T < 10$ eV
studied here. However, this is probably not so for Al at 1/4 of the
normal density, or for lower densities. We have less experience
with Au-plasmas to assess the quality of the resistivities for Au
obtaind here.
\subsection{Numerical results in the static limit.}
	The static resistivities for Al calculated using the above
methods (and some extensions of it)
 have been compared with experiment by Benage et al.\cite{benageR}
Given the uncertainties in the experiment and the various approximations
in the theory, the agreement is quite good.
However, there are a number of difficulties in the calculation. 
If we consider an Al-plasma at a compression $\kappa$=0.25, at $T> 2.5$ eV
the Al-ion has the bound shells $1s, 2s, 2p$, and $3s$. The $3s$ level
becomes increasingly shallow as the temperature is {\it reduced}. Below
approximately $T \sim 2.5$ eV, the $3s$ level begins to ``evaporate''
 and becomes
free, i.e, the $3s$ bound electrons ionize.
Even at 2.5 eV, the occupation of the $3s$ level is $\sim$  0.744.
 The ionization state $\bar{Z}$ increases from
$\sim 1.667$ to 3. The model where we have a single average $\bar{Z}$ is
clearly not satisfactory in such a region. The plasma contains
an equilibrium mixture of several ionization states $\bar{Z_i}$ with
concentrations $x_i$ such that
\begin{equation}
\bar{Z}=\Sigma_i x_i\bar{Z_i}
\end{equation}
For such situations (and indeed in general), the concentrations  $x_i$
have to be determined from the minimum of the total free energy. If 
$n$ species of ions are possible, then $n$ different
Kohn-Sham equations have to be solved, and $n(n+1)/2$ ion-ion
distribution functions have to be determined,  and the total
free energy has to be calculated as a function of $n$ and $x_i$.
The minimum property of the free energy yields the equilibrium
plasma conditions
from which the resistivity is calculated, using the $n$ scattering 
cross sections  and the $n(n+1)/2$ structure factors.
We reported such a calculation for $Al$
in Ref.~\cite{eos95}. Our experience is that, even with a multi-ion
fluid model, the self-consistent equations
 may fail to converge
since the $3s$ level (or the $2p$ level)
affects the iterative procedure. Majumdar and Kohn have
shown that physical properties of a system should be continuous across
the region where an electronic state moves from being a bound state
to a continuum state\cite{majum}. Hence we may calculate the
resistivity in two adjacent regions separated by a non-convergent
 region, and smoothly join the calculated resistivity 
across the ``difficult'' region. In Table~\ref{K-S} we show the Kohn-Sham
eigenvalues of the bound-electron states for Al ions in 
various
plasmas. Although the Kohn-Sham eigenvalues do not exactly
correspond to the excitation energies, they provide an initial
estimate which can be improved using the methods of time-dependent
density functional theory,
 or using self-energy calculations\cite{dyson-mott}.
$\,$
\begin{table}
\caption
{ Resistivity (in $\mu$ohm$\,$cm) of Al plsamsa as a function of the 
compression $\kappa$ and the temperature in eV, calculated within the
 approach described in the text.
 }
\begin{ruledtabular}
\begin{tabular}{lcccccc}

 T$\downarrow \;\;\; \kappa\to$& 0.1 & 0.25 & 0.5 & 1.0 & 2.0 & 4.0 \\
\hline\\
0.10  &1620 & 801 & 77   & 28   & 38   & 48    \\
0.25  &1582 & 812 & 84.1 & 30.5 & 37.4 & 47.0  \\
0.50  &1527 & 847 & 92.1 & 34.4 & 35.8 & 45.6  \\
0.75  &1447 & 859 & 99.7 & 36.7 & 34.5 & 45.0  \\
1.00  &1340 & 846 & 107  & 38.7 & 34.5 & 44.8  \\
1.75  &1063 & 707 & 122  & 44.4 & 34.2 & 44.6  \\
2.50  &844  & 597 & 135  & 50.2 & 35.4 & 44.7  \\
3.75  &1196 & 448 & 148  & 60.4 & 39.5 & 46.0  \\
5.00  &916  & 434 & 160  & 70.8 & 45.1 & 48.3  \\
7.50  &646  & 380 & 180  & 90.6 & 57.7 & 54.7  \\
10.0  &544  & 379 & 194  & 108. & 70.1 & 61.8  \\
\end{tabular}
\end{ruledtabular}
\label{restab}
\end{table}
%
%
\begin{figure}
\includegraphics*[width=6.5 cm, height=8.0 cm]{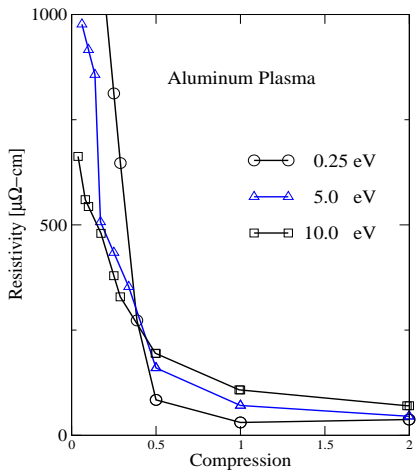}
\caption
{Resistivity of warm dense Aluminum as a function of
the compression for several temperatures
}
\label{fig1}
\end{figure}
\begin{table}
\caption
{
Kohn-Sham energy-level structure for several $Al$-plasmas within the
NPA
averag-configuration model }
\begin{ruledtabular}
\begin{tabular}{ccccccc}

Level &$\kappa=4$& $\kappa=2$& $\kappa=1$&$\kappa=0.5$&$\kappa=0.25$&$\kappa=0.1$\\
\hline\\

 &T=10  &&&&&\\
$\bar{Z}$& 3.043 & 3.0166 & 3.0164 & 3.0194 & 2.6060& 2.2427 \\
2s    &-5.7016  &-6.3615 &-6.9213 &-7.4089 &-7.9601&-8.5825 \\
2p    &-2.9311  &-3.5922 &-4.1529 &-4.6411 &-5.1928&-5.8159 \\
3s    & --      & --     &  --    & --     &-0.1871&-0.6168 \\
3p    & --      & --     &  --    & --     & --    &-0.1643 \\
&&&&&&\\
&T=2.5 &&&&&\\
$\bar{Z}$& 3.0427 & 3.0051 & 3.0003& 3.0000 & 1.6643& 1.6240 \\
2s      &-5.5247 &-6.1819 &-6.6655&-7.0257 &-7.4435& -7.6634 \\
2p      &-2.7838 &-3.4422 &-3.9268&-4.2875 &-4.7055&-4.9258 \\
3s    & --       & --     &  --   &  --    &-0.1178&-0.2758 \\
&&&&&&\\
&T=0.1 &&&&&\\
$\bar{Z}$& 3.0437 & 3.0053 & 3.0003& 3.0000 &does not &does not \\
2s      &-5.4903 &-6.1444 &-6.6282&-6.9811 &converge &converge \\
2p      &-2.7462 &-3.4008 &-3.8853&-4.2393 &due to     &due to \\
3s      &  --    &  --    &  --   &        & 3s level  &3s level \\
\end{tabular}
\end{ruledtabular}
\label{K-S}
\end{table}
As the density decreases and the temperature increases,
we inevitably pass through regions of $T$, $\kappa$ where
the problem of bound states which hover near ionization 
 becomes important.  We return to this question
in discussing the dynamic conductivity of Al-plasma
 at $\kappa$ = 0.25,
 and 0.1.
\begin{figure}
\includegraphics*[width=6.5 cm, height=8.0 cm]{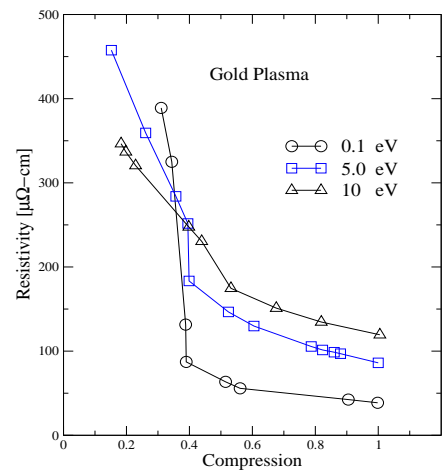}
\caption
{Resistivity of warm dense gold as a function of
the compression for several temperatures
}
\label{fig1au}
\end{figure}
Table I shows that, for $\kappa$ = 0.1 the resistivity essentially
decreases with $T$, while for  $\kappa$ = 0.25 the resistivity gradually
 increases in value, goes through a plateau-like region,
and then begins to decrease with temperature. 
The same behaviour holds for the other higher compressions, but the
table given here does not go high enough in temperature
 (for the higher densities) 
to show the
plateau effect and the onset of the decrease of $R$. Some authors
have interpretted the plateau in the resistivity as indicating a situation
where the mean-free path has become as small as possible ( as in the
Joffe-Regel rule). We {\it disagree} with this explanation of the
existence of a resistivity plateau in these cases in terms of a
saturation of the mean-free path. Quantitative agreement is
provided with a very {\it different picture}
 of what happens in the plasma. As the
material is heated, its resistivity increases just as in a metal, due to
the increased availability of a strip of states (of width $k_BT$)
at the Fermi surface
for the ``phonon-like'' scattering to take place. 
 However, as $k_BT$ increases, the number of
current carrying electrons also increases, compensating the
 resistivity increase.
During this process the chemical potential $\mu$  of the plasma electrons
 begins to decrease, and 
a temperature $T_{\mu=0}$ is reached when 
 $\mu$ passes through zero and towards negative values. Then the
Fermi sphere is gets broken down and from
 then on, {\it all} the electrons,
and not just those near the Fermi surface,
begin to conduct. The plasma is essentially classical. 
The resistivity {\it decreases}
 as the temperature increases.
We are in fact in the Spitzer-like regime. The plateau defines
the transition to the Spitzer-like regime.
The Fermi energy of the $\kappa$ = 0.1 case is small and
it is already behaving like a classical plasma.
 In  reality the picture is 
more complicated since the Fermi surface changes not only
 because of the temperature,
but also because the ionization of bound electrons changes
the value of $\bar{Z}$.
 This pushes the value of
 $T_{\mu=0}$, and the onset of the plateau to higher temperatures
 than in a model 
with constant $\bar{Z}$. It is easy to allow for this in a calculation of
 $T_{\mu=0}$ and confirm that the resistivity $R(T,\kappa)$
 given in our
table (and in Ref.\cite{milsch}) is consistent with this picture. 
\subsection{Contribution to the resistivity from electron-electron scattering.}
	Discussions of the electrical resistivity of
 plasmas sometimes contain
allusions to the e-e contribution to the electrical resistivity.
 However, the
electron-current operator $\vec{j}=(e/m)\vec{k}$ commutes with the
 electron-electron interaction
Hamiltonian $H_{ee}$. 
\begin{equation}
\vec{j}H_{ee}-H_{ee}\vec{j}=0
\end{equation}
This shows that the current is  {\it conserved}
under the e-e interaction.
Hence electron-electron interactions {\it cannot} contribute to
the resistivity arising from the electron current.
However, the e-e interaction has an indirect
effect since it screens the electron-ion pseudopotential
 $V_{e-i}(q)$. That is, electron-ion vertices must occur
in all diagrams which contribute to the resistivity.
In perturbation-theory
 approaches to the conductivity or resistivity,
 it is quite easy to get a contribution
 to $R$ from e-e scattering alone, if the theory is carried
 out only to, say, second order.
This means, if an all-oder calculation were done, the higher
 order corrections would exactly
cancel the low-oder result for the e-e scattering contribution.
Electron-electron interactions contribute to the resistivity of
solids where the periodic potential generates {\it Umkapp} scattering.
But this is not the case in plasmas if they can be considered
uniform to within a length scale significantly 
larger than the mean free path. Classical transport calculations
for systems with gradients (i.e, no translational invariance)
is well known\cite{matt}. The quantum calculation is also
well known in transport across heterostructures\cite{lanbut}. 
\section{Dynamic conductivity.}
	If we apply a field $\vec{E}\, cos(\omega t)$ to the system,
say
using a light probe of frequency $\omega$, then the polarization of the
medium is described by the polarization function $\Pi(q,\omega)$ which is
directly connected with the transverse dielectric function
 $\varepsilon(\omega)$ and the dynamic conductivity $\sigma(\omega)$. The 
wavevector $q$ of the photon is nearly zero
and may usually be omitted. 
 The real part $\sigma_r$=$\Re\sigma(\omega)$ at $\omega\to 0$ reduces to the
static conductivity $\sigma$ that was already discussed.
\begin{eqnarray}
\varepsilon(\omega)&=&1-\frac{\omega^2_p}{\omega^2}-
\frac{4\pi}{\omega^2}\Pi(\omega) \\
\sigma(\omega)&=&i\frac{\omega^2_p}{4\pi\omega^2}
+i\frac{\Pi(\omega)}{\omega}
\label{sigmapi}
\end{eqnarray}
	If the effect of interactions and bound states is small, 
the conductivity of "free" electrons driven by the field
$E_0exp(i\omega t)$, and damped by scatterng  at ion centers
 is well approximated 
 by the Drude model. It uses a relaxation time $\tau_0$, (or a
damping parameter $\gamma_0$)
independent of the frequency. In partially ionized systems, or when there are
interband effects in solids, it is necessary to include bound-free
 and bound-bond contributions to $\sigma(\omega)$. A useful practical form is
 the extension of the Drude model where a model dielectric function is used.
Thus,
\begin{eqnarray}
\label{modeleps}
\varepsilon(\omega)&=&\varepsilon_r+i\varepsilon_i \\
  &=&1-\frac{\omega^2_{p0}}{\omega(\omega+i\gamma_0)}
-\sum^{\lambda_{m}}_{\lambda=1}\frac{\omega^2_{p\lambda}}
{\omega(\omega+i\gamma_{\lambda})-\epsilon_{\lambda}}\\
\sigma_r(\omega)&=&\frac{\omega\varepsilon_i(\omega)}{4\pi}
\end{eqnarray}
Here the pure free-electron Drude term uses the damping parameter
 $\gamma_0$ and a plasma frequency $\omega_{p0}$, while other processes
are modeled by a finite set of oscillators
 parametrized by $\gamma_{\lambda}$ and
$\omega_{p\lambda}$. This is a form useful for fitting
experimental data since reflection and transmission
 experiments could be used to
extract best fit values of $\gamma_{\lambda}$,$\omega_{p\lambda}$,
 subject to the sum rule
(f-sum rule):
\begin{equation}
\int_0^\infty d\omega \sigma_r(\omega)=
\frac{1}{4}\sum^{\lambda_m}_{\lambda=0}\omega_{p\lambda}=\pi n_{tot} e^2/m_e
\end{equation}
Here $n_{tot}$ involves the free electrons
 (given by the average charge $\bar{Z}$)
as well as the electrons occupying the localized
 atomic states participating in the
optical transitions. Unlike in liquid-metal studies, these are not readily 
available for warm dense partially degenerate plasmas. Further, the electron
populations in the bound and free states are linked by the Fermi distribution
associated with the electron temperature and the chemical potential.
 Hence this LTE
(local thermodynmaic equilibrium) type constraint also should be imposed in fitting
the experimental data to the model dielectric function.
 Thus the most fruitful approach
is to use the values of the mean ionization $\bar{Z}$, occupation numbers,
electron chemical potential
 $\mu_e$, $\gamma_{\lambda}$, $\epsilon_{\lambda}$ etc.,
provided from the density-functional NPA calculation
 in setting up the fitting process.
The simple Drude equation with a {\it fixed} value of $\gamma_0$ satisfies the
f-sum rule.
It can be shown that if we define $\Pi(\omega)$ in terms of the
non-interacting polarizability $\Pi^0(\omega)$, by the
form:
\begin{equation}
\Pi(\omega)=\Pi^0(\omega)
/[1+G(\omega)\Pi^0(\omega)]
\end{equation}
where $G(\omega)$ is called a ``local-field'' correction, then the sum rule
is satisfied if it holds for $\Pi^0(\omega)$. Thus, if
$\Pi(\omega)$ can be calculated, then we can 
calculate the dynamic conductivity from it.

 The effect of corrections to the simple Drude term
arising from interband terms in
 solid Aluminum, were examined many years ago by Dresselhaus, Harrison, and
by Ashcroft and Sturm.\cite{ash-sturm}. They showed that there are
band to band transitions near $\sim$.05 and 0.15 of
 the Fermi energy $E_F$ (which is
$\sim 12$ eV. for normal density $Al$). The optical conductivity of
normal-density liquid Al
slightly above  the melting point has been measured by Miller\cite{miller} and 
shows a simple Drude form. A discussion of liquid metal data including
Cu, Ag and Au is found in Faber\cite{faber}.
The deviations from the Drude form seen in solid-Al arise from the
 splitting of degenerate bands due to the crystal potential, and to
normal interband transitions. Benedict et al. have argued that the
thermal broadening of the electron self-energy is by itself sufficient to
``wash out'' these solid-state effects, even if the ion lattice remained
intact\cite{benedict}, as may be the case in  short-pulse laser
generated plasmas. The effect of electron-electron as well as ion-electron
interactions on the electron self-energy at finite-$T$ was also discussed by
us for the hydrogenic case\cite{dyson-mott}

   When we consider
warm dense $Al$-plasmas, the Kohn-Sham eigenvalues
 can be used to assess if the
driving field   $\omega$ could excite bond-bound or bound-free
processes. As seen from Table~\ref{K-S}, the $2s$ levels for the
compressed systems ($\kappa$ =0.1, 0.5, 1, 2, 4) occur
 between -7.4  au to -5.5 au.,
while the $2p$ level ranges from -4.6 au. to -2.7 au. Hence, the Drude
formula would be reasonable for these  systems and for standard
optical probes. The situation is quite different for 
the $\kappa$ =0.25 case. Here 
 the $2p$ level ranges from $\sim$  -4.5 au. to -5.5 au., but the
$3s$ level is very close to the ionization threshold. At $T$ = 8 eV, the
$3s$ level is at -0.165 au, and rises to -0.105 au. near $T$= 3 eV, and
then completely disappears for $T$ below $\sim$ ~2.5 eV. 
Since a 300-400 nm optical probes corresponds to about 4-3 eV,
 it is clear that
such probes would show deviations from the Drude conductivity for the
 $\kappa$ =0.25 case. When we go to low temperatures, the $\bar{z}=3$
ionization is very stable, while the $3s$ level creates the presence of
$\bar{z}=1,2$ 
ionization states. In the case of $\kappa$ = 0.1, we have {\it two} 
shallow states , viz., $3s$ and $3p$. The conductivity derived from
the model dielectric function, i.e., Eq.~\ref{modeleps} would be relevant to
the representation of experimental data for such systems. In the following
we look at first-principles calculations.
\subsection{Perturbational approach to the dynamic conductivity.}
	In a theory of $\sigma(\omega)$ 
we are in effect looking for the
polarization function $\Pi(\omega)$. 
If the potential 
 $V_{ie}(q)$ is weak,
(no bound states) diagrammatic methods can be used. However,
the second-order expression in the screened interaction is essentially the
only one that is tractable for realistic potentials.
A number of
such quasi-second order results exist in the literature, derived using various methods.
An old result, due to Hopfield, holds if the structure factor is
essentially unity\cite{hopfield}.
\begin{equation}
\Re\sigma(\omega)=-\frac{\omega_p^2}{(4\pi\omega)^2}
\int \frac{d\vec{q}}{(2\pi)^3}q^2q_z|V_{ie}(q)|^2\Im[\frac{1}{\varepsilon(q,\omega)}]
\end{equation}
A more complete result, including the contribution from the
dynamic structure factor of the ions can be written down, using
an approach similar to that given by Mahan for phonons.\cite{mahan}
R\u{o}pke et al. have also discussed second-order expressions
 within the Zubarev approach which is most suitable for short-ranged
potentials free of bound states.\cite{ropke} The approach
 used by Mahan is more interesting and may be used for 
long-range potentials.
In fact, even in the two-temperature case where the ions are at a
temperature $T_i$, while the electrons are at a temperature $T_e$,
 it is easy to show  using the Keldysh technique
 (assuming that the probe
frequency $\omega$ does not overlap with core transitions)
that the frequency dependent collisional relaxation time $\tau(\omega)$
is\cite{milsch},
\begin{eqnarray}
\label{cond}
\lefteqn{\tau^{-1}(\omega)=-(\omega^2_p\omega)^{-1}\int\frac{q^2dqM^2_q}{(2\pi)^3}
q_z^2\int\frac{d\nu}{2\pi}}\cdot   \\
& &\Im[\chi_{ii}(q,\nu)]\Im[\chi_{ee}(q,\nu+\omega)]
\Delta N(T_e,T_i,\nu,\omega) \nonumber
\end{eqnarray}
where we have set
\begin{equation}
\Delta N(T_e,T_i,\nu,\omega)=[N\{\beta_e(\nu+\omega)\}
-N\{\beta_i(\nu)\}]. 
\end{equation}
Here $N\{\beta_i(\nu)\}$ is a Bose factor giving the occupation number
of density-fluctuation modes at the temperature $T_i$ and energy $\nu$.
For the  equilibrium situation we simply set $T_i=T_e$.
Then, as $\omega\to 0$, this equation can be shown to reduce to the
inverse collision time used in the Ziman formula.
In Eq.~\ref{cond} the electron response $\chi_{ee}$
 and the ion-response $\chi_{ii}$
 mediate the  energy and momentum exchange between the two
subsystems. The
imaginary parts of the response functions are related to the
dynamic structure factors by a relation of the form:
\begin{equation}
S(k,\omega)=-\frac{1}
{2\pi}\coth\big[\frac{\omega}{2T}\big]\Im\chi(k,\omega)
\end{equation}
In the case of the electrons, $\chi(k,\omega)$ can be written down in
terms of the Lindhard function $\chi^0(k,\omega)$ and the local-field
correction $G(k,\omega)$. In the limit where the electrons becomes
classical, the Lindhard function simply reduces to the Vaslov
function.
The dynamic structure factor of the ions can also be modeled
in a similar fashion.\cite{elr} Clearly, if the
probe frequency $\omega$ is smaller than the electron plasma frequency,
then we may make a static approximation for  $G(k,\omega)$. 
Since normal-density Al (and even Au, depending on the compression)
 have  high plasma frequencies, this is
a good approximation for dense Al or Au. However,
the static $G(k,\omega=0)$ used should be such
that at least the compressibility sum rule is satisfied.
Also, we can introduce an $\omega$ dependent model
$G(k,\omega)$ with $G(k,\omega)$= $G(k,\omega=0)$ for $\omega<\omega_P$,
and $G(k,\omega)$ fitted to the high-frequency moment sum rules
for $\omega>\omega_P$. In practice it is not known how to
satisfy all the
sum rules.
The static pair-distribution function $g(r)$
 recovered from
 the 
imaginary part of such a response function should also agree with the
known $g(r)$ of the electrons at that density. The successful calculation
of the $g(r)$, $S(k)$, and the $G(k)$ in a consistent manner for
 electrons at strong-coupling and arbitrary temperatures (i.e, arbitrary
 degeneracies) was presented 
recently in Ref.~\cite{chncT} 

Although an extension of the above equation to take  account
of bound-free and bound-bound electron processes can be written down
``by hand'', it is not easy to provide a systematic development.
More fruitful approaches are to use density-functional molecular-dynamics
simulations or
 time-dependent
density functional theory. We consider these below.
\subsection{Conductivity via the Kubo-Greenwood formula and molecular dynamics.}
\begin{figure}
\includegraphics*[width=8.5 cm, height=11.0 cm]{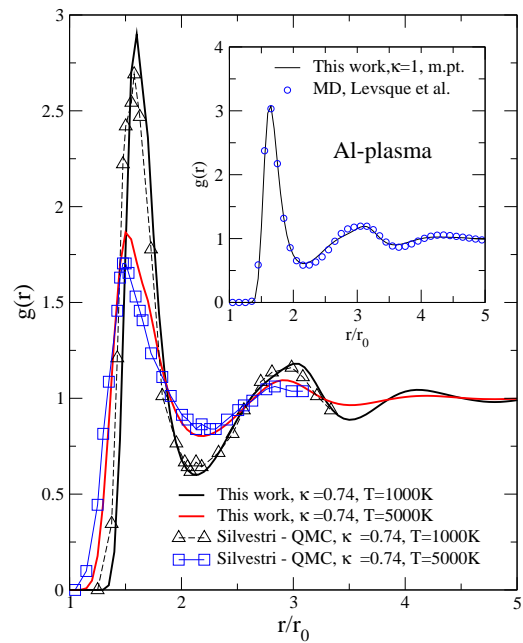}
\caption
{Ion-ion pair-distribution functions of Al from this work compared with
those obtained by Silvestri\cite{silv1}. The inset shows the comparison at
normal density and at the melting point. Here no data from
ref.\cite{silv1} is availble, but we use accurate molecular-dynamics
simulations (Levesque et al\cite{lwr}) for comparison.
}
\label{figgr}
\end{figure}

Another approach to the conductivity $\sigma(\omega)$ is to 
use molecular dynamics to develop the ionic-liquid structure,
 while retaining
a DFT approach only for the electronic structure\cite{silv1,alavi,gillan}.
 The
 ion subsystem is modeled with, say, typically
  32-256 atoms in a simulation box
of volume $V_b$ which is periodically repeated. 
An externally constructed pseudopotential is used and an ionization model is
{\it assumed}. The
required number of electrons based on the ionization model
is placed in the box.  The ions are held at some fixed ionic
configuration \{$R_i$\} and the Kohn-Sham electronic 
wavefunctions $\psi_f(r,\{R_i\})$ and energies $\epsilon_f$ are computed.
Given the size of the system, it is not practical to do more
than a few k-points; usually only one $k$-point, e.g.,
the
$\Gamma$ point is computed. The conductivity for the given configuration,
and for the selected $k$ point with weight $W(k,\{R_i\})$
is estimated using the second-order
Fermi golden rule formula.
\begin{eqnarray*}
\sigma(\omega,\{R_i\},k)&=&\frac{2\pi}{3V_b\omega}
\sum_{fg}|W(R_i, k)<\psi_f|\vec{p}|\psi_g>|^2\\
& &(n_f(k)-n_g(k))
\delta(\epsilon_f(k)-\epsilon_g(k)-\omega)  
\end{eqnarray*}
This is in fact the Kubo-Greenwood (KG) procedure (sampled with
one k-point) that one may use for
 a crystalline
solid. A Kohn-Sham form is used instead of the
many-electron eigenstates. The occupation factors $n_f(k)$ are also
the one-particle occupations at the appropriate temperature.
The energies used are the LDA eigenvalues without corrections.
 Even a crystalline
structure has its phonon modes and their effect is ignored in the
energy denominators.
Similarly, the self-energy effects from the electron-electron
processes are also ignored, although they can be quite
large\cite{dyson-mott,benedict}.
Only the {\it umklapp} processes associated with the reciprocal vectors of
the simulation box contribute to the static conductivity.
 This is because the 
wavefunctions are eigenstates of the ``crystalline'' structure \{$R_i$\}. The
actual conductivity has to be obtained by taking a configuration average which
requires the Helhmoltz free energies for all crystal configurations \{$R_i$\}.
In effect, the method uses simple LDA-DFT for the electrons,
 but abstains from using
DFT for the ions and carries out a detailed MD evaluation
 of the liquid structure.
We believe that this is unnecessary, especially for systems where the
spherical symmetry of the plasma is a statistically reasonable assumption.
The work of Kown et al.\cite{kwon} on strongly-coupled H-plasmas using
these methods, and their comparison with our work is an example of this.
 In Fig.~\ref{figgr} we compare the
ion-ion pair-distribution functions obtained from our NPA+
HNC+bridge type procedures with available simulations
from Silvestrelli et al\cite{silv1}, and from Levesque et al\cite{lwr}.
A simulation box of 125 atoms implies only 5 atoms per dimension and hence
even the central atom feels only two atomic shells around it.
The calculation of the $\omega\to 0$ limit needed to obtain a static
conductivity is also
quite difficult, and one approach is to
 {\it assume the validity of the Drude form} and
 fit a free-electron Drude form to achieve this.
The Kubo-Greenwod type MD procedure would nevertheless
provide useful complementary results for
comparison with our two-component DFT approaches, and would be of
much interest in studying low-temperature systems with a tendency
to covalent bonding  and clustering. 
 However,
these methods bring in a number of approximations of their own and
need cautious reconsideration. In this context, an interesting test
would be to examine expanded liquid-Hg\cite{bhat}
 using the Kubo-Greenwood MD approach. 
 \begin{figure}
\includegraphics*[width=8.5cm, height=11.0cm]{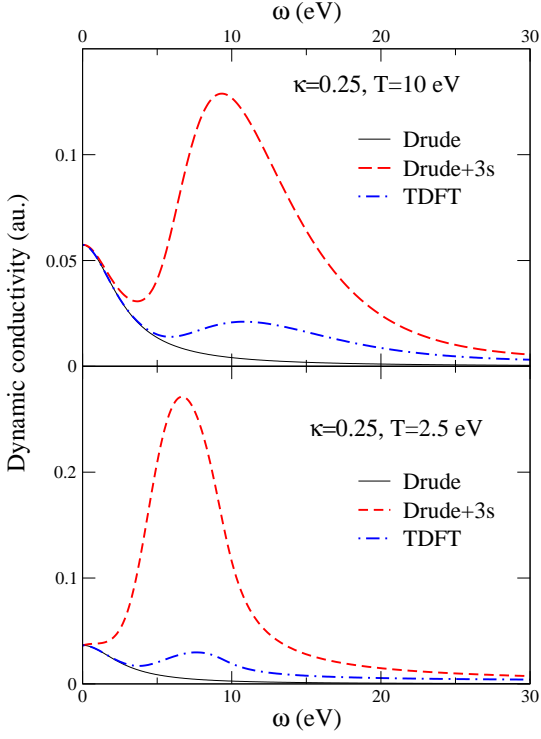}
\caption
{Dynamic conductivity (au.) of Al at a compression
of 0.25 at $T$ =2.5 eV and 10 eV. The Drude conductivity
is modified by the presence of the $3s$ level. Time-dependent
DFT calculation greatly weakens the effect.
}
\label{kp25}
\end{figure}
\subsection{ Dynamic conductivity from
 time-dependent density functional theory.}
Time dependent DFT provides a convenient approach
to the calculation of the dynamic conductivity of interacting systems.
The TDFT formulation relevant to dense plasmas\cite{gri}
includes dynamic screening and coupling to ion-dynamics
in a computationally convenient, self-consistent manner.
The main consequence of the TDFT formulation is to replace, say,
the dipole matix elements $<i|\vec{r}|j>$ between states $i,j$ by a
dynamic form $<i|\vec{r}(\omega)|j>$ where the modification
of the driving field by the response of the system is taken into
account in a self-consistent manner.
Consider a weak external field $\vec{E}_{ext}(t)$ = $\vec{E}cos(\omega t)$.
This corresponds to an external potential:
\begin{equation}
U_{ext}(\vec{r},t)= e\vec{r}\cdot E_{ext}(t)
\end{equation}
The dipole form of the interaction is used since one of the objectives is to
include the corrections  arising from the
presence of bound states. However, the dipole form of the matrix elements
can be easily replaced by the momentum or acceleration formulation
 when needed. We assume that the
electric field is directed along the z-direction,
 and suppress vector notation for the
field unless needed.
The external potential induces an electron density fluctuation $\delta n(r)$
which in turn generates corrections to the Coulomb and exchange-correlation
potentials. Since the linear absorption coefficient (or optical conductivity)
is the object of our study, $\delta n(r)$ etc., can be written in terms of the 
electron response function $\chi^0(r,r'|\omega)$ which
can be approxmiately constructed\cite{gri} from the 
Kohn-Sham eigenstates of the plasma. Then we have 
\begin{eqnarray}
\label{uind}
U(\vec{r},\omega)&=& U_{ext}(\vec{r},\omega)+V^c_{ind}+V^{xc}_{ind}\\
V^c_{ind}&=&\int d\vec{r'}
\frac{\delta n(\vec{r},\omega)}{|\vec{r}-\vec{r'}|}\\
V^{xc}_{ind}&=&\left[\frac{\partial V_{xc}(\vec{r},\omega)}
{\partial n(\vec{r},\omega)}\right]\delta n(\vec{r},\omega)\\
\delta n(\vec{r},\omega)&=&
\int  d\vec{r'}\chi(\vec{r},\vec{r'}|\omega)U(\vec{r'},\omega)
\end{eqnarray}
Here $V^{xc}_{ind}(r)$ is calculated from the
 gradient $\partial V_{xc}/\partial n$
evaluated at the density $n_0(r)$ in the unperturbed neutral pseudo atom.
 Since the form of
the time-dependent exchange-correlation potential $V_{xc}(r,\omega)$
is still not established, most implementations
 use the 
static  $V_{xc}[n(r)]$
of ordinary density functional theory.
The above set of equations have to be solved self consistently to obtain the
total perturbing potential $U(\vec{r},\omega)$. In effect, the dipole operator,
or equivalently, the momentum operator of the scattering
 electron is replaced by
a space and time dependent quantity which enters into the polarizability.
Given the spherical symmetry of the system, the total perturbing 
potential has the form
\begin{equation}
U(\vec{r},\omega)=-(1/2)EU_{\omega}(r)Y^0_1(\vec{r}/r)
\end{equation}
Here $Y^0_1$ is a spherical harmonic.
If we ignored these induced fields, the conductivity of
 the system can be written as:
\begin{equation}
\Re\,\sigma^0(\omega)= \frac{e^2}{a_0\hbar}
\Im\,\sum_{\nu,\nu'}\frac{\pi N_i \omega}{3}
\frac{|<\nu|r|\nu'>|^2(f_\nu-f_{nu'})}
{\omega+\epsilon_\nu-\epsilon_\nu'+i\delta}
\end{equation}
Here all the quantities on the right of the summation are in atomic units.
The $e^2/(a_0\hbar)$ factor is the atomic unit of
 conductivity.
 $N_i$ is the number of ions per unit volume,
 and we neglect the effect of ion-ion
correlations in the bound-free and bound-bound processes.
(It can be shown that these contribute mainly to the width
 of the transition by broadening the levels)
Also when $\nu,\nu'$ =
$k,l,m$ and $k',l'm'$ for free-free transitions,
 the dipole matrix element is replaced by the
momentum form, i.e, $|<\nu|\vec{r}|\nu'>|^2$ =
 $|<\nu|\vec{\nabla}|\nu'>|^2/\omega^2$.

The conductivity expression 
$\sigma^0(\omega)$ is known to be particularly inadequate when treating
b-f and b-b processes, unless the initial bound state is a deep
lying  (e.g, 1s) state\cite{kedge}.  Hence we need to include the
induced fields 
in treating under-dense plasmas where
there is a bound state close to ionization.
This involves the use of the total perturbing
potential, $U(\vec{r},\omega)$, rather than the external
potential $\vec{r}\cdot\vec{E}$ in constructing the
matrix elements contained in the conductivity expression: 
\begin{eqnarray}
\lefteqn{\Re\,\sigma(\omega)=\frac{ N_i\pi \omega}{3}\Im\sum_{\nu,\nu'}}\\
& &|<\nu|U_{\omega}(r)Y^0_1(\vec{r}/r)|\nu'>|^2
 *(f_\nu-f_{nu'})
\delta(\omega+\epsilon_{\nu}-k^2/2) \nonumber
\end{eqnarray}
The effect of level broadening can be included in the above
expression by replacing the delta function $\delta(\omega+\epsilon_{\nu}-k^2/2)$
by a form containing the self-energy corrections to the single-particle levels,
as in Grimaldi et al, where self-energies were calculated. However, for
our present purpose, we use the $\tau$ used in the Drude calculation
to provide a broadening parameter.
In fact, we have shown in Ref.~\cite{dyson-mott} that the self-energy
contribution of ion-electron scattering to level broadening is
identical to that given by the Ziman formula.
 Thus, setting $\gamma=1/\tau$, we
replace the $\delta$-function in the above equation with a Lorentzian.
$$\delta(\omega+\epsilon_{\nu}-k^2/2)\to \frac{\gamma/\pi}
{(\omega+\epsilon_{\nu}-k^2/2)^2+\gamma^2}$$
\subsection{Some numerical results.}
	In this section we present results for the dynamic conductivity of 
some Al-plasmas to illustrate the effect of the scattering events where
shallow bound states modify the Drude-like conductivity. These
``bound states'' are really Kohn-Sham eigenvalues and hence their
values may need improvement by evaluating the corrections
using a Dyson equation.\cite{dyson-mott}
The first step of the calculation is the solution of the NPA model to
obtain the Kohn-Sham basis set $\{\phi(r)_\nu\}$ for
 the given electron density
and temperature. The codes necessary for the NPA calculation,
the resistivity calculation, as
well as the plasma conditions (EOS) that go into the resistivity
or conductivity calculations may be accessed via the internet
at our website\cite{web}.
\begin{figure}
\includegraphics*[width=8.5cm, height=11.0cm]{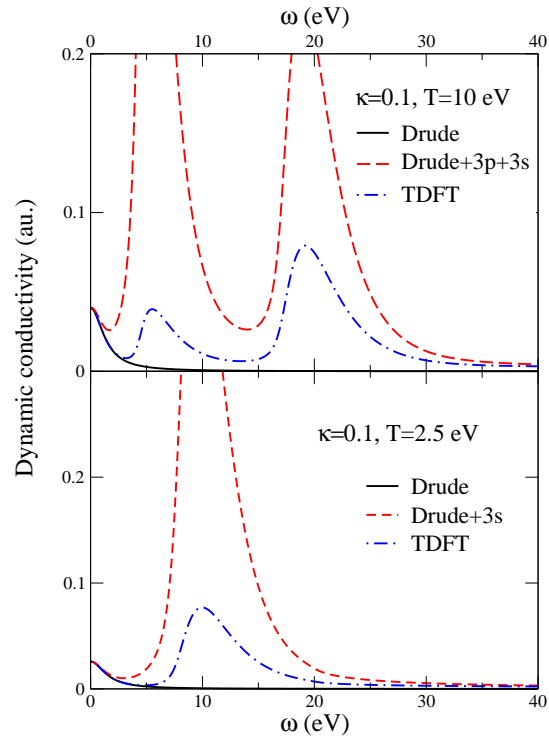}
\caption
{
Dynamic conductivity (au.) of Al at a compression
of 0.1 at $T$  =2.5 eV and 10 eV. The Drude conductivity
is modified by the presence of shallow $3s$ at $T$ = 2.5 eV, and
also the  $3p$ level at 10 eV. The effect is reduced in the
TDFT calculation.
}
\label{kp1}
\end{figure}
%
%
Some results for the shallow bound states present in
$Al$-plasmas at T=2.5 eV and 10 eV, for compressions $\kappa=0.25$, and 0.1
are given in table II. In the T=10, $\kappa=0.1$ case we have two shallow
bound states, viz., $3s$ and $3p$. 
In Table~\ref{occ} their occupation numbers and the interacting
chemical potential needed for calculating the Fermi factors are given.
In calculating the matrix element $|<nlm|r|kl'm'>|^2$, the
correct density of states $\aleph(\epsilon)$ 
for the continuum state $|kl'm'>$, normalized
in a sphere of radius $R$, with $R\to\infty$ should be 
included.
That is,
\begin{eqnarray*}
k_nR-\pi l/2&=&n\pi +\delta_{kl}\\
 \aleph(\epsilon)&=&\frac{\partial n}{\partial k}
\frac{\partial k}{\partial \epsilon}
\end{eqnarray*}
The calculations are very simple if (a)
we ignore the Zangwill-Soven type TDFT effects arising from the
reaction of the system which act to  modify the external field.
(b) if we ignore the phase shifts $\delta_{kl}$,
and replace the boundstates by hydrogen-like states with the correct
$Z=\bar{Z}$.
The final results depend on the level broadening parameter
$\gamma$ used in the calculation.

In fig.\ref{kp25}  and Fig.\ref{kp1} we show the conductivity
 arising from 
the presence of the $3s$ and $3p$ levels, as well as the Drude term, and the
modification when TDFT is included in an approximate manner.
These TDFT calculations should be regarded as highly provisional
results only; in fact, the treatment of shallow boundstates using
DFT is itself questionable since the Kohn-Sham eigenvalues
are known to be a poor approximation to the actual excitation spectrum.
\subsubsection{bound-bound processes.}
Clearly, the presence of a partially occupied  $3s$ and a $3p$ state 
with an energy separation of the order of 0.46 au. would lead to a
contribution from b-b processes at around 12 eV. This contribution is easily
included in the calculation under the simplifying assumptions that we
noted before. However, this is a  relatively sharp ``line'' resonance 
which is expected to undergoes
significant modification when the time-dependent effects are taken in to
account. We have not included it in
our figures.
\begin{table}
\caption{details regarding the n=3 shallow states in
underdense $Al$ at T=10 eV and 2.5 eV. Full occupation is when Occ
is unity}
\begin{ruledtabular}
\begin{tabular}{cccccc}
$\kappa$&0.25 & 0.25&0.1& 0.1 \\
$ T$ &2.5 eV & 10 eV &2.5 eV & 10 eV\\
\hline\\
$\mu$  &0.0393&-0.5538&-0.6652  &-0.9589\\
Occ(3s)&0.7444&,0.2217& 0.6850  &0.1455 \\
Occ(3p)& --   &  --   &  --     &0.0843\\
\end{tabular}
\end{ruledtabular}
\label{occ}
\end{table}
\section{Conclusion.}
 The first-principles calculation of the
dynamic conductivity of warm dense matter may be conveniently
carried out within the
framework of multi-component density functional theory.
 The static calculation
(NPA etc) provides the Kohn-Sham basis set, phase shifts, pseudopotentials
for constructing ion-ion pair potentials and structure factors. These
immediately provide results for the static conductivity. Further, the
energy-levels and occupation numbers obtained from the Kohn-Sham
NPA solution can be the starting approximation for a time-dependent
density functional calculation of the optical conductivity. This
proceeds in much the same way as the optical absorption cross section.
Details of such a time-dependent calculation, based on the
method of Zangwill and Soven,  may be found in
Grimaldi, Lecourte and Dharma-wardana.\cite{gri}


\end{document}